# AN INTERSTATE TRIPS ANALYSIS DURING COVID-19 IN THE UNITED STATES

By leveraging the mobile phone location data collected from over 100 million anonymous devices, this study identified interstate trips and analyzed the temporal changes and spatial differences of interstate mobility behavior during the COVID-19 in the United States.

**Weiyi Zhou[1], Minha Lee[1], Qianqian Sun[1], Weiyu Luo[1], Chenfeng Xiong[1], Lei Zhang[1,*]**

[1]Maryland Transportation Institute, Department of Civil and Environment Engineering
University of Maryland
1173 Glenn Martin Hall, College Park, MD 20742

Corresponding Author:

**Lei Zhang, Ph.D.**
Herbert Rabin Distinguished Professor
Director, Maryland Transportation Institute
Department of Civil and Environment Engineering
University of Maryland
1173 Glenn Martin Hall, College Park, MD 20742
Email: lei@umd.edu
Phone: +1 (301)405-2881

**Abstract**

The worldwide outbreak of COVID-19 has posed a dire threat to the public. Human mobility has changed in various ways over the course of the pandemic. Despite current studies on common mobility metrics, research specifically on state-to-state mobility is very limited. By leveraging the mobile phone location data from over 100 million anonymous devices, we estimate the population flow between all states in the United States. We first analyze the temporal pattern and spatial differences of between-state flow from January 1, 2020 to May 15, 2020. Then, with repeated measures ANOVA and post-hoc analysis, we discern different time-course patterns of between-state population flow by pandemic severity groups. A further analysis shows moderate to high correlation between the flow reduction and the pandemic severity, the strength of which varies with different policies. This paper is promising in predicting imported cases.

**Introduction**

Since December, 2019, the novel coronavirus disease, COVID-19, has been spreading globally at an alarming speed. On March 11 2020, the coronavirus outbreak was officially declared a global pandemic by the World Health Organization. At the time, there were 118,319 confirmed cases worldwide and 114 countries or territories affected (*1*). Immediately after this upgraded warning of a global health emergency, the U.S. government issued a national emergency on March 13 and launched coronavirus guidelines for America on March 16, in which the social distancing intervention was suggested to curb the pandemic (*2-5*). On March 19, California became the first state in the U.S. to announce a statewide stay-at-home order. Over the following three weeks, more than forty state governments successively promulgated similar orders throughout or partially throughout their respective states. After a period of nationwide social distancing intervention, Alaska, Georgia, Oklahoma, and South Carolina became the first four states to partially reopen on April 24. By May 20, all 50 states had partially reopened.

The outbreak poses a serious threat to society. To fight against this coronavirus, it is critical for the public to follow social distancing orders. Various mobility metrics of population flow are closely associated with the spread of COVID-19 (*6*). Various common mobility metrics, such as percentage of people staying home, miles traveled per person, and number of trips per person, are thoroughly analyzed in a previous paper by the authors (*7*). External trips (defined as trips occurring between two different analysis zones) particularly provide a perspective of the population flows between different regions (*8*). External trip data often act as a critical component to project the number of cases for regional spread in epidemic modeling as well as imported cases in epidemic modeling (*6, 9, 10*). In the epidemiologic aspect, analyzing the external trips enlightens researchers and policymakers about the effect of social distancing intervention. From a social science point of view, it allows us to understand human mobility behavior in response to the pandemic and the mobility intervention (*7, 11-13*). Additionally, understanding the external trip change and the influence of other factors supports decision-makers in determining next steps for both control measures and society reopening.

As revealed from the literature, external trips play an important role during the pandemic. However, there is a lack of research on the reality of external trips in the United States. Our paper particularly focuses on external trips in the United States at the state level; that is, interstate trips based on real-world mobility data. This could act as a great source to understand the disease spread as well as public reactions toward the urgent threat (*14*). Mobile device location data, a promising data source for analyzing mobility behaviors, have recently gained more attention by researchers and commercial corporations (*15, 16*).

During the current era of COVID-19, mobile data have been actively utilized to explore pandemic-related behaviors and prevent the spread of the virus (*17, 18*).

By leveraging the valuable information of real-world mobile location data, the authors developed a set of advanced and peer-reviewed algorithms of identifying trips and weighting to population (*19*). Then the daily interstate trips for the entire United States from January 1, 2020 to May 28, 2020 (weekends and holidays are removed from analysis) are extracted and aggregated from the results using our previously developed algorithms. This study investigates the patterns of interstate trips in relation to the pandemic through the following aspects. First, a spatiotemporal analysis of 50 states (and the District of Columbia), including pre-pandemic and the different phases of pandemic, is conducted to demonstrate the interstate trip variations. Second, the study explores the relationship between the reduction of interstate trips and the disease severity. Third, the high-, middle-, and low-severity states are compared to show the time course changes of interstate trips through Repeated Measures ANOVA and post-hoc analysis.

## Results
   1. Spatiotemporal Pattern of Interstate Trips

Fig. 1 presents the temporal changes of inflow interstate trips by state and the spatial differences. Fig. 1A shows the absolute scale of weekdays' average (no holidays) attracted interstate trips during the baseline period (from January 2 to January 31) while others (Fig. 1B, C, D, E, F) indicate the percentage changes compared to the baseline during selected weeks. In January, states differ much regarding inflow interstate trips. During the week of February 28 and the week of March 13, nearly all states show a percentage increase, which is consistent with the natural travel pattern of increased traffic during March. Due to the outbreak of COVID-19, such an increasing trend is terminated and a significant nationwide decrease is observed, as shown during the week of March 27 and the week of April 10. Afterwards, with the partial reopening of society, the inflow interstate trips increase again and some states even exceed the baseline level in the most recent week.

We also conduct a more detailed analysis of interstate trips at the origin-destination level between states, as shown in Fig. 2, with states in both axis ordered by the cumulative cases by April 6 (when all states announced stay-in-home orders) in ascending sequence. The absolute trip volume between states for four periods are displayed in the top plots (Fig. 2A, B, C, D), while the bottom ones (Fig. 2E, F, G, H) correspondingly present the interstate trip changes between states. During the week of the national emergency announcement, the interstate trips share a similar spatial trend as the baseline with slight increase (Fig. 2A). Those states with more confirmed cases also have higher external trips between each other, which is also observed in three other periods (Fig. 2E, F, G, H). It intuitively reveals the correlation between human mobility and virus spread. With states successively announcing stay-at-home orders, the demand for travel begins to decline, especially for interstate trips. Around 40% of nationwide decline can be observed during the week when all states issued stay-at-home orders (Fig. 2B). Such decline is more obvious in those states with more cases as shown by clustered blue dots in Fig. 2B, such as New York, Louisiana, and New Jersey. During the week that partial reopening started in four states, an overall rebound between states can be observed. Those states with higher pandemic severity still show a more notable decline (Fig.

2C). When all 50 states started partial reopening, a more obvious increase is presented and 62% of state-to-state external trips exceed the baseline level (Fig. 2D). Yet, some reopening states remain at a decreased status, such as Florida, Texas, and Georgia, which are relatively more severely infected (Fig. 2H). It is noteworthy that the daily average of new cases is still reported as many as 21,770 in the U.S. during the reopening stage. Although the interstate trips continue to increase amid COVID-19, it might be hasty to determine that the impact of the pandemic on the external trips is weakened. The relationship between the two is further analyzed in the next section.

2. **Relationship Between Interstate Trips Reduction and Outbreak Severity**

Based on the spatiotemporal changes presented in the section above, it is hypothesized that the percentage change of external trips is sensitive to the severity of COVID-19. The daily Spearman correlation coefficient between percentage change of inflow interstate trips and cumulative cases per 1,000 people (defined as the indicator of pandemic severity) of all 50 states and the District of Columbia shows interesting time-course changes (Fig. 3). This temporal change also generally coincides with that of national new cases. Considering that external trips at the state level are mainly long-distance trips that are probably planned ahead, a seven-day time lag is applied when doing correlation analysis since interstate trips might not respond instantaneously to the severity of coronavirus. Obviously, the inflow external trips are negatively influenced by COVID-19, which means that the more severe the pandemic is in a state, the less external trips it attracts. The degree of such influence, additionally, varies over different pandemic phases. In the early stage of the outbreak, beginning with the national emergency declaration on March 13, there is a decreasing trend of Spearman coefficient, reaching the lowest bound of -0.537, which indicates a notable negative correlation. Moreover, this downward trend (i.e., increasing strength of negative correlation) is more stable after March 19 when California first announced their stay-at-home order. The period suggests a notable signal of behavior change that the pandemic severity in the destination state is taken more seriously by travelers. Subsequently, the correlation coefficient is stabilized with mild fluctuations from April 6, when all states announced stay-at-home orders, to April 24, when the first state announced reopening. During this period, the public experiences an inertia in status without further decline of interstate trips. This could be the ultimate state of social distancing and there are mainly unavoidable interstate trips. Afterwards, an increasing trend of negative correlation is shown following April 24, indicating that the impact of COVID-19 on interstate trips gradually decreases. With the reopening of more and more states, there is a reduced impact of the pandemic on interstate trips. After May 20 with all states already reopening, the trend flattens out but still presents a moderate negative correlation. This may infer a reduced attention to the pandemic under the influence of reopening. Nevertheless, such attention remains at a moderate level as shown by the negative Spearman coefficient around -0.3.

3. **Repeated Measures One-way ANOVA and Post-hoc Analysis**

With repeated measures analysis, a time course pattern of the dependent variable of interest can be established showing response time points with significant effect and durations (*20, 21*). In a one-way Repeated Measures ANOVA (RM-ANOVA), the same group of subjects are measured under different levels determined by a single factor, such as at multiple time points, so that the individual differences within the same group do not account for the post treatment measures (*22*). This advantage of eliminating the individual differences yields a smaller error variability than a simple ANOVA, which is more sensitive to the different levels (*21, 23, 24*). The RM-ANOVA can be applied to a non-experimental design involving natural changes over time (*25*). Upon a significant time effect shown from RM-ANOVA, post hoc tests are expected, which identify the critical time points indicating a significant difference from the others (*25*).

With repeated measures one-way ANOVA and post hoc analysis, we investigate the course of interstate trip variations by pandemic severity groups. The study separates 50 states (and the District of Columbia) into groups of three based on the severity of pandemic, i.e., cumulative confirmed cases on April 6. The statistical results for high-, mid-, and low- severity groups are shown in Fig. 4 respectively, which reveal several phenomena, as follows. First, the inflow interstate trips of states in the high-severity group experience a rapid decline phase from March 16 to March 23. This is indicated by a complete independent pink square (pink means being insignificant) in Fig. 4A. In contrast, states of the low-severity group do not have such a particular phase (Fig. 4C) and states of the mid-severity group are somewhere in between (Fig. 4B). This suggests that the interstate trips towards the states with a higher pandemic severity decrease at a quicker speed. Trips to another state are sensitive to the pandemic situation at destination. In the late stage of the studied period, especially after April 26, more insignificant cells (colored in pink) are associated with earlier dates, which infers that there is a rebound phenomenon of inflow interstate trips. Those insignificant cells occur more frequently and earlier in states with lower pandemic severity. The low-severity group rebounds first, the high-severity group rebounds last, and the mid-severity group is again somewhere in between. Overall, the analysis shows that, as far as the percentage change of interstate trips is concerned, the public reaction to the pandemic is fairly sensitive to the situation at the destination area and it varies during different phases of the pandemic.

**Discussion**
The most obvious finding to emerge from this study is that people's mobility change in terms of interstate trips between states is influenced by the progression of the pandemic. In spite of a nationwide decline after the national emergency, the response to the pandemic presents a spatiotemporal difference. In the early stage of the pandemic, the public presents a quick travel behavior change with a notable percentage decrease of interstate trips. Later on, such decline slows down and gradually levels off. What is more interesting is that the OD-level change of interstate trips is fairly sensitive to the pandemic situation at trips destinations. Those states with a relatively higher pandemic severity generally present a more significant reduction of attracted interstate trips. This discriminative treatment of destination states reflects the public perception of risk and precaution of avoiding the exposure to the virus. Furthermore, the degree of this discrimination varies over time. At the beginning, these differential interstate trip reductions of states are increasingly correlated with the pandemic prevalence at those destination states. This correlation reaches the peak with a Spearman score of -0.5 until most of the states issued stay-at-home orders, and remains at a stable level until the beginning of society reopening. As more and more states are reopened, the degree of this discrimination first weakens and then gradually stabilizes with a Spearman score of -0.3. In order to further investigate this discrimination phenomenon, all states are grouped by high-, mid-, and low-severity. The within-group analysis and group-wise comparisons suggest that the high-severity group of states has a unique stage of decline at a statistically significant higher speed than others. In addition, the attracted interstate trips by the low-severity group of states rebound earlier than the others.

In spite of its limitations, this study certainly offers some insight into the interstate trip pattern under the influence of a pandemic. The results, meanwhile, suggest the importance of considering public policies and pandemic severity in the short-term prediction of interstate trips. The findings can additionally be utilized to forecast the imported cases at a state level.

**Materials and Methods**
  1. **Trip Identification**

Raw mobile location data are the sitings (latitude and longitude at a timestamp) for each device. In order to estimate aggregated mobility between different regions, identifying the individual trips from raw location data is the first step of data processing. A rule-based algorithm with three thresholds (i.e. time threshold, distance threshold, and speed threshold) is developed and utilized for trip identification. Among the three thresholds, the speed threshold indicates the movement of points, while the distance and time thresholds are utilized to determine the termination of the movement. Each point has three possible states: belonging to an existing trip, belonging to a new trip, and invalid record. The state of belonging to an existing trip is assigned when the point satisfies several conditions determined by speed, time, and distance thresholds. Otherwise, either the point will be identified as starting a new trip if the speed from this point to the next point exceeds the speed threshold, or the point is treated as an invalid record when the speed is below the threshold.

  2. **Weighting**

Despite the large sample size, the identified trips are still a sample of the population. In order to make the results represent population-level statistics of travel behavior, we applied both device-level and trip-level weighting. The device-level weighting uses a market penetration method, which expands the observed sample devices to population level. The weights are calculated by the ratio of total population to the observed residents' devices in each county. Take a county with 500 total population for example; if 200 observed devices are determined to reside in this county, a device-level weight of 50 is assigned to the 200 devices. The trip-level weights, additionally, are applied. Since the trajectories for each device may not be continuously observed, we develop trip-level weights as makeup factors. First, we derived trip rates (i.e., number of daily trips per person) from the National Household Travel Survey (NHTS) for all states. Then for each state, the ratio of trip rate from NHTS to the weekdays' average trip rate of the first two weeks of February from our data is treated as trip-level weight. For example, if NHTS trip rate for a given state is 3 and our data indicates 2, then a trip-level weight of 1.5 is applied to the devices in this state.

3. **Inter-State Trips**

For each identified trip, the trip origin and destination (latitude and longitude) are spatially joined with the state shape file. Accordingly, each trip is represented as originating from a state to another. Inter-state trips are an aggregated number of trips occurring between any two given states from day to day. Weekends and federal holidays, including January 1 (New Year's Day), January 20 (Birthday of Martin Luther King Jr), and February 17 (President's Day), are excluded to eliminate noise.

**Author Contributions**
WZ, QS, CX and LZ designed the study. WZ, QS, and ML analyzed the data. WZ, QS, and ML interpreted the data. WZ, QS, and ML wrote the first draft. All authors contributed to the final draft.

**Competing Interests**
The authors declare no competing interests.

**Data Availability**
WZ is responsible for addressing any correspondence and material requests.

**Acknowledgments**
We would like to thank and acknowledge our partners and data sources in this effort: (1) Amazon Web Services and its Senior Solutions Architect, Jianjun Xu, for providing cloud computing and technical support; (2) computational algorithms developed and validated in a previous USDOT Federal Highway Administration's Exploratory Advanced Research Program project; and (3) COVID-19 confirmed case data from the Johns Hopkins University Github repository and sociodemographic data from the U.S. Census Bureau.

**Figures and Tables**

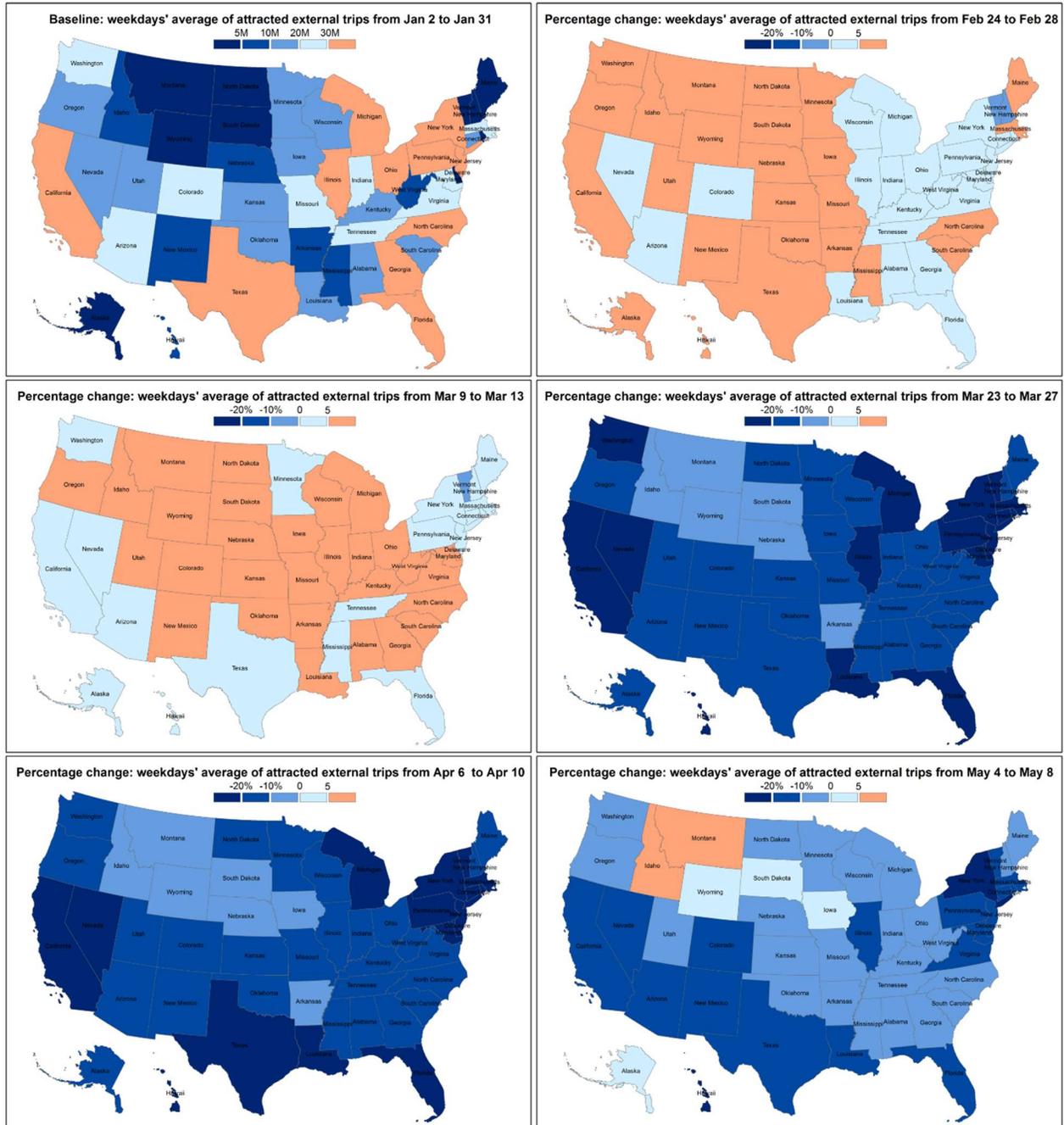

**Fig. 1 Spatiotemporal percentage change of interstate trips attracted by states.** (**A**) The weekdays' average of interstate trips from January 2 to January 31 shows that different states have different baselines. (**B**) (**C**) The percentage change of attracted interstate trips before the national emergency during the week ending on February 28 (B) and on March 13 (C), compared with the baseline show a general nationwide increase. (**D**) (**E**) The percentage change during the second (D) and the fourth (E) week after the national emergency compared with the baseline show a significant nationwide decline. (**F**) The most recent week shows an obvious nationwide increase with partial reopening started in most states.

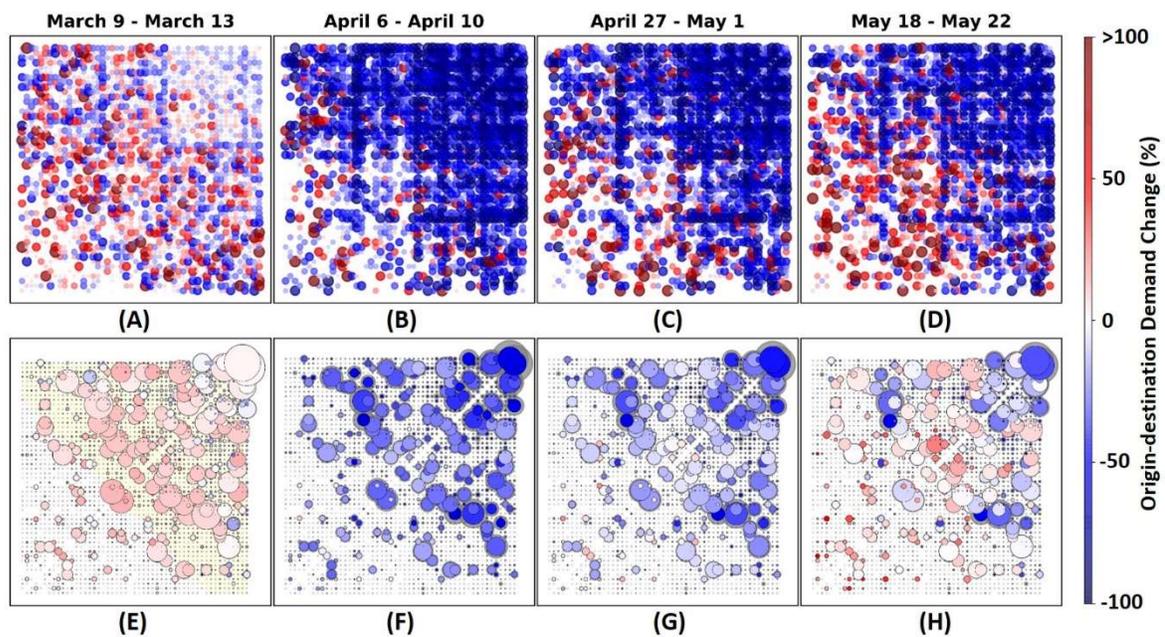

**Fig. 2 The percentage change of interstate trips relative to baseline and corresponding absolute scale of interstate trips.** The baseline is weekdays' average (no holidays) of January. Both x and y axis are the states ordered by cumulative cases as of April 6 in ascending sequence and hence each dot corresponds to each state pair. The color in all subplots indicates the percentage change and the dot size in the bottom four subplots shows the absolute scale of interstate trips. (**A**) (**B**) (**C**) (**D**) are the percentage change of weekdays' average during the four selected weeks compared to the baseline. (**E**) (**F**) (**G**) (**H**) additionally show the absolute scale for each state pair.

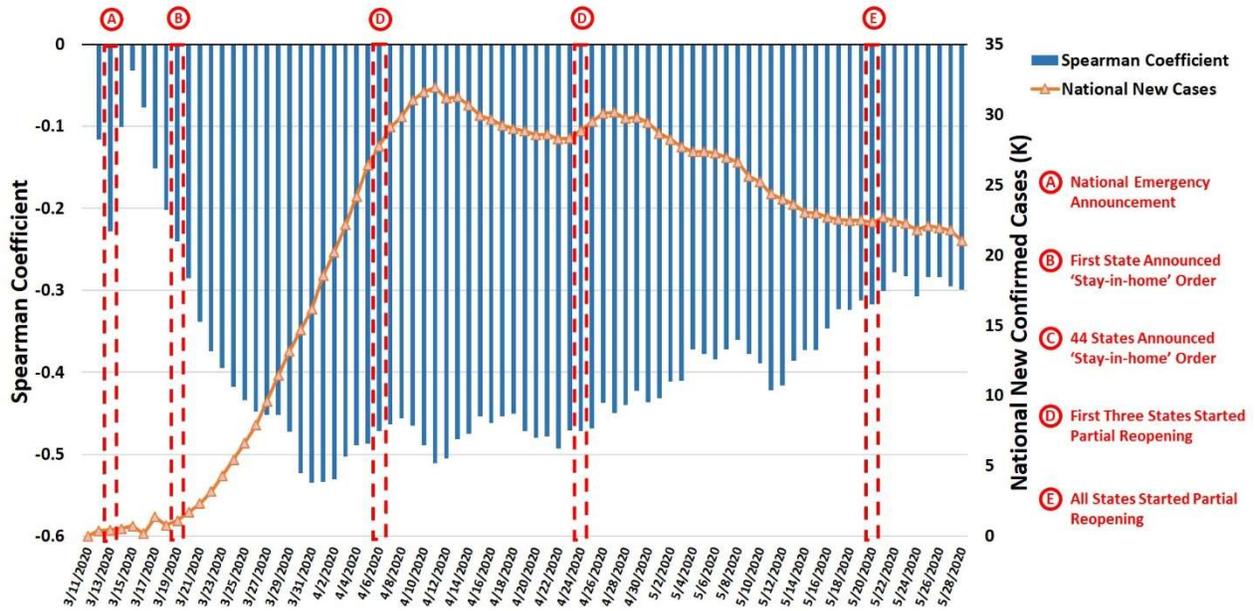

**Fig. 3 Daily Spearman correlation coefficient between percentage change of inflow interstate trips and cumulative cases along with the national daily new cases.** A seven-day time lag is applied to the inflow interstate trips. X axis shows the dates of cumulative cases. Five critical time points are annotated: March 13 (national emergency), March 19 (California first announced stay-at-home order), April 6 (all states announced stay-in-home order), April 24 (first three states partially reopened), and May 20 (all states partially reopened). The temporal dynamics of the correlation present interesting changes amid the COVID-19 cases and different policies.

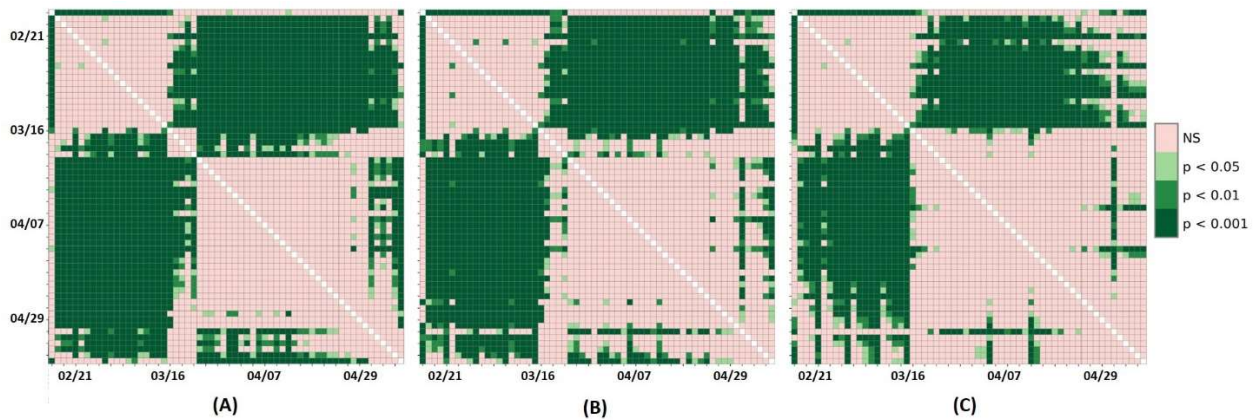

**Fig. 4 Significance plots of inflow interstate trips for different severity groups of states.** The significance plot is a matrix, in which each cell shows the p-value of testing two days on x and y axis, respectively. All states are grouped into three severity levels. The testing result for each group is shown by **A** (high-severity group), **B** (mid-severity group), and **C** (low-severity group). The plots indicate different time-course patterns for each group.